\documentclass[preprint]{aastex6}

\usepackage{graphicx}			
\usepackage{amssymb}				
\usepackage{braket}
\usepackage{amsmath}
\usepackage{rotating}
\usepackage{float}
\usepackage{tabularx}

\fullcollaborationName{The Friends of AASTeX Collaboration}

\begin{document}


\title{Rate constants for fine-structure excitations in O - H collisions with error bars obtained by machine learning}


\author{Daniel Vieira\altaffilmark{1,2} and Roman V. Krems\altaffilmark{1}}
\affil{Department of Chemistry, University of British Columbia \\
Vancouver, BC V6T 1Z1, Canada}

\begin{abstract}
We present an approach using a combination of coupled channel scattering calculations with a machine-learning technique based on Gaussian Process regression to determine the sensitivity of the rate constants for non-adiabatic transitions in inelastic atomic collisions 
to variations of the underlying adiabatic interaction potentials.
Using this approach, we improve the previous computations of the rate constants for the fine-structure transitions in collisions of O($^3P_j$) with atomic H. 
We compute the error bars of the rate constants corresponding to 20 \% variations of the ab initio potentials and show that this method can be used to determine which of the individual adiabatic potentials are more or less important for the outcome of different fine-structure changing collisions. 
\end{abstract}

\section{Introduction} \label{sec:intro}

Of the elements heavier than helium, oxygen is the most abundant in the interstellar medium. Interstellar oxygen is predominantly in the form O \footnotesize I \normalsize as it is shielded from ionizing radiation by the more abundant hydrogen atoms. The fine-structure transitions in collisions of O \footnotesize I \normalsize with atomic hydrogen 
\begin{equation}
\mathrm{O}(^{3}P_{j}) + \mathrm{H}(^2S) \rightarrow \mathrm{O}(^{3}P_{j'}) + \mathrm{H}(^2S)
\label{reaction}
\end{equation}
are hence an important cooling process in the interstellar medium (Dalgarno \& McCray 1972; Cowie \& Songaila 1986, Pequignot 1990, Shaw et al. 2006, Glover \& Clark 2013) and may be a dominant mechanism at low temperatures (Dalgarno \& McCray 1972). 

The relative intensities of the O \footnotesize I \normalsize spectral lines serve as an important diagnostic probe of the atomic environment (Cowie \& Songaila 1986, Jenkins \& Tripp 2011). The absorption lines can be used as a source of information on the column densities for the $^{3}P_{j}$ levels, and the strong emission lines in diffuse regions can be used to study a diverse range of nebulae (Pequignot 1990).
Accurate rate constants for fine-structure transitions (\ref{reaction}) are needed for understanding cooling in early galaxies (Greif et al. 2010), atomic-line diagnostics and gas modelling in protoplanetary disks (Meijerink et al. 2008, Tilling et al. 2012) and the spectroscopy of stellar atmospheres (Fabbian et al. 2009). The rates of the reaction (\ref{reaction}) have been used in models of turbulence in the interstellar medium (Lesaffre et al. 2013, Guillard et al. 2015) and studies of both the ionized and neutral parts of the warm interstellar medium (Dong \& Draine 2011, Jenkins 2013). 

Most of these models rely on the theoretical calculations of collision-induced {\color{black} fine-structure} transition rates by Launay \& Roueff (1977) and, more recently, by Abrahamsson et al (2007). The interaction of O($^{3}$$P$) with H is governed by four molecular potentials. 
Abrahamsson et al (2007) used the interaction potentials computed by Parlant \& Yarkony (1999). However, the potentials used by Abrahamsson et al (2007) were originally constructed for the scattering calculations of the collision rates for the electronic excitations O($^{3}$$P$) $\rightarrow$ O($^{1}$$D$). As we will show below, adapting the potentials to the collision problem (\ref{reaction}) resulted in an error, rendering the calculations of Abrahamsson et al (2007) inaccurate. In addition, the interaction potentials of Parlant \& Yarkony (1999), as any intermolecular potentials, are subject to uncertainties associated with errors of the ab initio calculations. It is important to evaluate the effect of these uncertainties on the rate constants. However, because the collision problem is parametrized by four potentials and it is not known which of these potentials play a dominant role in determining the {\color{black} fine-structure} transitions, estimating the errors of the rate constants stemming from uncertainties in the potentials presents a challenge. 

In the present paper, we develop an approach using a combination of coupled channel scattering calculations with a machine-learning technique based on Gaussian Process regression to determine the sensitivity of the rate constants to variations of the four interaction potentials.
Using this approach, we repeat the calculations of Abrahamsson et al (2007) to report accurate values of the rate constants based on the potentials of Parlant \& Yarkony (1999) and compute the error bars of the rate constants corresponding to reasonable variation of the ab initio potentials. We show that this method can be used to determine the relative importance of the individual potentials in determining the outcome of different {\color{black} fine-structure} changing collisions.

\section{Computation Details}

\begin{figure}
\centering
\includegraphics[scale=0.3]{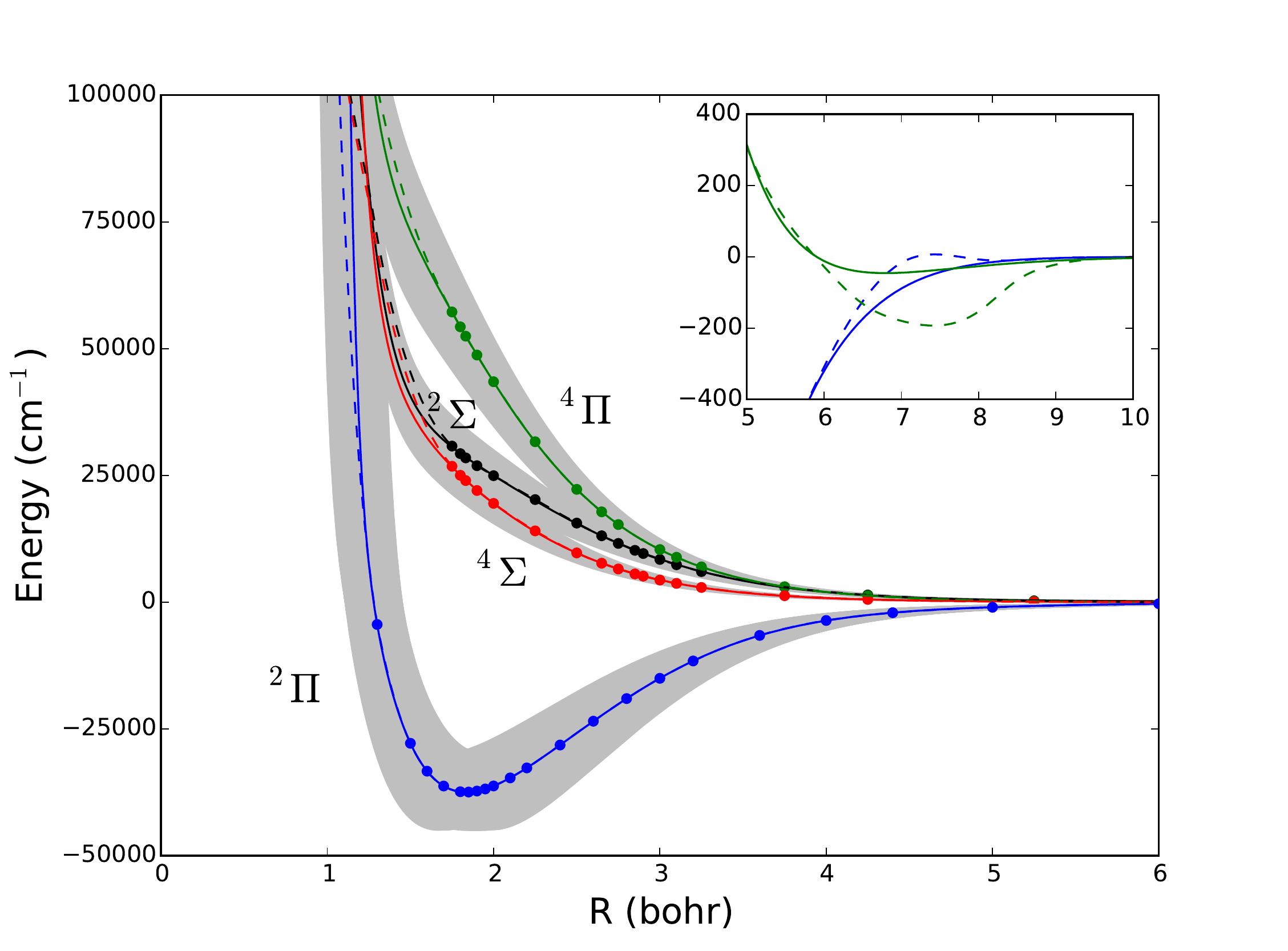}
\caption{The interaction potentials for the X$^{2}$$\Pi$, $^{2}$$\Sigma$, $^{4}$$\Pi$ and $^{4}$$\Sigma$ electronic states of OH that parametrize the collision problem (\ref{reaction}). 
The symbols represent the ab initio calculations by Parlant \& Yarkony (1999); the solid curves -- the analytical fits of the ab initio results;  the dashed curves -- the potentials used by Abrahamsson et al (2007). 
The grey bands show the variation of the interaction potentials considered in the present work for determining the sensitivity of the rate constants to uncertainties in the molecular potentials. The inset compares the analytical fits of the ab initio data used in the present work with the potentials used by Abrahamsson et al (2007). 
}
\end{figure}

The collisions of O($^3$$P$) with H($^2$$S$) are determined by four interaction potentials of the following symmetries: X$^{2}$$\Pi$, $^{2}$$\Sigma$, $^{4}$$\Pi$ and $^{4}$$\Sigma$. These potentials have been computed most recently by  
Parlant and Yarkony (1999). Figure 1 shows the calculations of Parlant and Yarkony (1999) as symbols. Instead of using the fixed interaction potentials for the quantum scattering calculations as was done by all previous authors, in the present work we 
ask the following question: if each of the potentials is uncertain to within a grey band shown in Figure 1, what is the corresponding uncertainty in the collision rates? 

This question could be answered by a quantum scattering calculation on a grid of parameters determining the variation of each of the four potentials. However, the number of such scattering calculations would have to be unfeasibly large. 
For example, if the variation of each potential was determined by $n$ parameters, the rate constants would be $4n$-dimensional functions, requiring a total of $\sim 10^{4n}$ scattering calculations of collision rates at each temperature in order to determine the global response of rates 
to the variation of the potentials. 
Instead, we propose a machine-learning approach based on training a statistical learning method with the results of quantum scattering calculations, reducing the required number of computations to a linear function of the number of parameters.

In order to account for the variation of the potentials, we first produce the analytical fits of the numerical data of Parlant and Yarkony (1999) shown in Figure 1 as solid curves. We then modulate each of the analytical fits of the repulsive potentials for the excited molecular states $^{2}$$\Sigma$, $^{4}$$\Pi$ and $^{4}$$\Sigma$ by a scaling factor varied from 0.8 to 1.2. The interaction potential of the ground state X$^{2}$$\Pi$ is modulated by means of two parameters: the overall scaling factor inducing a 20 \% change in the magnitude of the potential and a factor translating the interatomic distance by $\pm 10$ \%, leading to shifts in the position of the potential minimum, a change in the slope of the repulsive part of the interaction potential and a change of the classical turning point in the limit of zero collision energy. We thus parametrize the collision problem by five parameters ${\bf x} = \{ x_1, ..., x_5 \}$.

We use Latin Hypercube Sampling (LHS) scheme to generate $Q$ sets of these parameters at random (Stein 1987, McKay et al. 1979). LHS ensures that these sets sample the five-dimensional space evenly. 
For each set of the five parameters ${\bf x}$, we compute the scattering rates for the {\color{black} fine-structure} transitions (\ref{reaction}). The computations are performed using the rigorous coupled channel 
method described in multiple previous papers, including the work by Launay \& Roueff (1977) and Launay (1977).  The rate constants are obtained by integrating the energy dependence of the scattering cross sections with the Maxwell-Boltzmann distribution. 
The cross sections for a fixed collision energy are computed by integrating a system of coupled channel equations  presented by Launay (1977). As recommended by NIST \footnote[1]{See https://www.nist.gov/pml/atomic-spectra-database}, we adapt the energies 158 cm$^{-1}$ for {\color{black} O($^3P_{j=1}$)}
 and 227 cm$^{-1}$ for O($^3P_{j=0}$) relative to the energy of O($^3P_{j=2}$). 

 The coupled channel calculations produce $Q$ values of the rate constants corresponding to $Q$ sets of the 
parameters ${\bf x}$ at each temperature. We treat each {\color{black} fine-structure} transition separately and compute only the rate constants for the collision-induced excitations: {\color{black} $j=2 \rightarrow j' = 1$}, 
{\color{black} $j=2 \rightarrow j' = 0$} and {\color{black} $j=1 \rightarrow j' = 0$}. These rate constants are used as input for Gaussian Process (GP) regression. 

GP regression is an efficient, statistical learning technique for interpolation in multi-dimensional spaces (Stein 1999, Cressie 1993, Quinonero-Candela \& Rasmussen 2005, Neal 1998, Rasmussen \& Williams 2006, Williams 1998, Papritz \& Stein 1999). 
The application of GP regressions to molecular physics problems has been described in our recent work (Cui \& Krems 2015, Cui, Li \& Krems 2015, Cui \& Krems 2016), where more details on the implementation of the method can be found. 
In brief, the interpolation starts by estimating the statistical correlations between rate constants corresponding to different points $\bf x$ in the five-dimensional space considered here. 
We approximate the correlation function as 
\begin{eqnarray}
R({\bf x},{\bf x}') = \left\{ \prod_{i=1}^{q = 5} \bigg(1+\frac{\sqrt{5}|x_i-x_i'|}{\omega_i}+\frac{5(x_i-x_i')^2}{3\omega_i^2}\bigg)\mathrm{exp}\bigg(-\frac{\sqrt{5}|x_i-x_i'|}{\omega_i}\bigg) \right\},
\label{correlation}
\end{eqnarray}
where $\omega_i$ are the unknown parameters representing the characteristic length scales of the correlations.
Note that the specific choice of the correlation function in Eq. (\ref{correlation}) is not unique.  
The mathematical form of the correlation function determines the efficiency of the GP regression, i.e. the number of rate constant calculations needed to obtain the global five-dimensional dependence on the parameters.
However, given any physically reasonable mathematical form of $R$, GP regression should become accurate, when based on a large enough number of rate constant calculations.

For a set of $Q$ points in the five-dimensional space, we construct the correlation matrix $\bf A$ whose elements are given by $R({\bf x},{\bf x}')$, with $\bf x$ and $\bf x'$ covering all pairs of rate constants in the computed set. 
To find the estimates of the parameters $\boldsymbol{\omega}=(\omega_1,\omega_2,\cdots,\omega_5)^\top$ we maximize the log-likelihood function
\begin{eqnarray}
\textrm{log}\mathcal{L}(\boldsymbol{\omega}|{\bf Y}) = -\frac{1}{2}\left[Q \textrm{log}\hat{\sigma}^2+ \textrm{log}(\textrm{det}(\textbf{A})) + Q \right],
\label{log-l}
\end{eqnarray}
by iteratively computing the determinant and the inverse of the matrix $\bf A$. 
In Eq. (\ref{log-l}),
\begin{eqnarray}
\hat{\sigma}^2 (\boldsymbol{\omega}) = \frac{1}{Q}(\textbf{\textit{Y}} -\boldsymbol{\beta})^{\top}{\bf{A}}^{-1}(\textbf{\textit{Y}} -\boldsymbol{\beta}),
\end{eqnarray} 
\begin{eqnarray}
{\hat{\beta}} (\boldsymbol{\omega}) = (\mathbf{I}^{\top}\textbf{A}^{-1}\mathbf{I})^{-1}\mathbf{I}^{\top}\textbf{A}^{-1}\textbf{\textit{Y}},
\end{eqnarray}
$\bf Y$ is the vector of all computed rate constants for a given {\color{black} fine-structure} transition, $\boldsymbol{\beta} = \beta {\bf I}$, $\bf I$ is the identity vector, 
and the hat over the symbol denotes the maximum likelihood estimators (i.e. expressions giving the maximum likelihood estimates).

\begin{figure}
\centering
\includegraphics[scale=0.5]{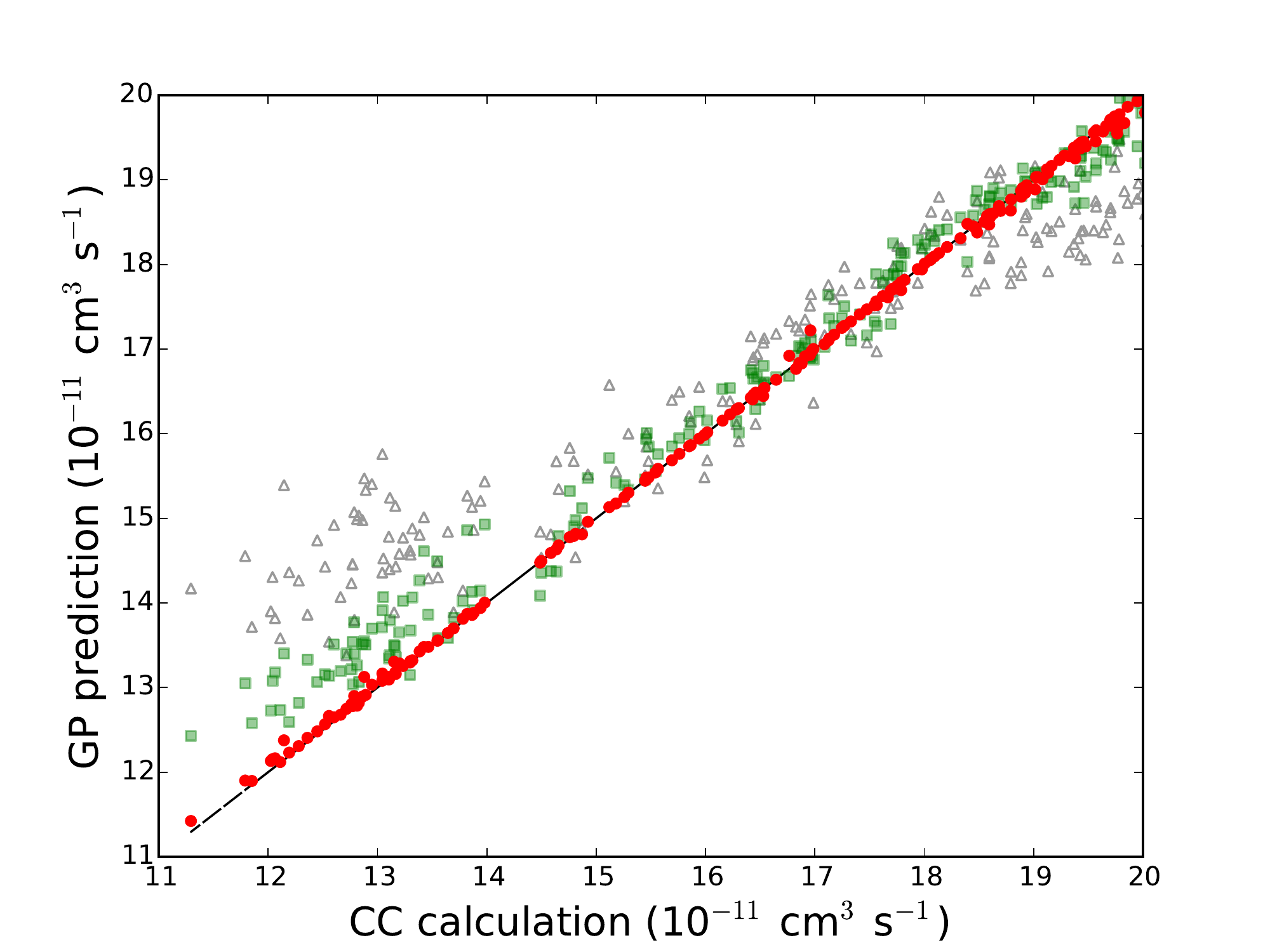}
\caption{Comparison of the rate constants from the coupled channel (CC) computations with the rate constants produced by the GP regression based on the different number of points: $Q=10$ (triangles), $Q=100$ (squares), $Q=1000$ (red circles).
The plot represents a sample of randomly selected 200 points in the five-dimensional parameter space that have not been used for GP regression. The calculations are for the {\color{black} $j = 1 \rightarrow j' = 0$} transitions at the temperature $T = 300$ K. The deviation from the diagonal line illustrates the error of the GP regression. 
}
\end{figure}

Given the best values of $\boldsymbol{\omega}=(\omega_1,\omega_2,\cdots,\omega_q)^\top$ thus found, we can construct the final correlation matrix 
$\bf A$. The prediction of the rate constant at any point $\bf x_0$ of the five-dimensional space $\bf x$ can then be expressed in terms of this matrix and a vector 
\begin{eqnarray}
 \mathbf{A}_0=
 \left[
\begin{array}{c}
R({\bf x}_0,{\bf x}_1)  \\
R({\bf x}_0,{\bf x}_2) \\
\vdots \\
R({\bf x}_0,{\bf x}_Q)
\end{array}
\right ]
  \end{eqnarray}
as follows: \begin{eqnarray}
\label{mean-prediction}
k({\bf x}_0)&=& {\beta}+\mathbf{A}_0^{\top}\mathbf{A}^{-1}({\bf Y} -\boldsymbol{\beta} ).  
\end{eqnarray} 
The vector $\bf A_0$ describes the statistical relationship between the rate constant at the point of interest $\bf x_0$ and all other computed rate constants assembled in the vector 
${\bf Y}$. Eq. (\ref{mean-prediction}) gives the five-dimensional dependence of the rates on the parameters modulating the OH interaction potentials. 

Given Eq. (\ref{mean-prediction}), we can compute the rates corresponding to the original potentials of Parlant \& Yarkony (1999), 
calculate the error bars arising from the variation of the interaction potentials and examine the sensitivity of the rates to each of the five different parameters. 
To do that, we apply variance-based sensitivity analysis (Mcrae et al. 1982, Saltelli et al. 1999, Saltelli \& Bolado 1998, Cacuci \& Ionescu-Bujor 2004, Helton \& Davis 2002). As illustrated in the following section, 
this analysis can be a powerful tool to explore the mechanisms of inelastic scattering by providing information  on which parts of the underlying interaction potentials are predominantly responsible for the collisional energy transfer.

\begin{figure}
\centering
\includegraphics[scale=0.4]{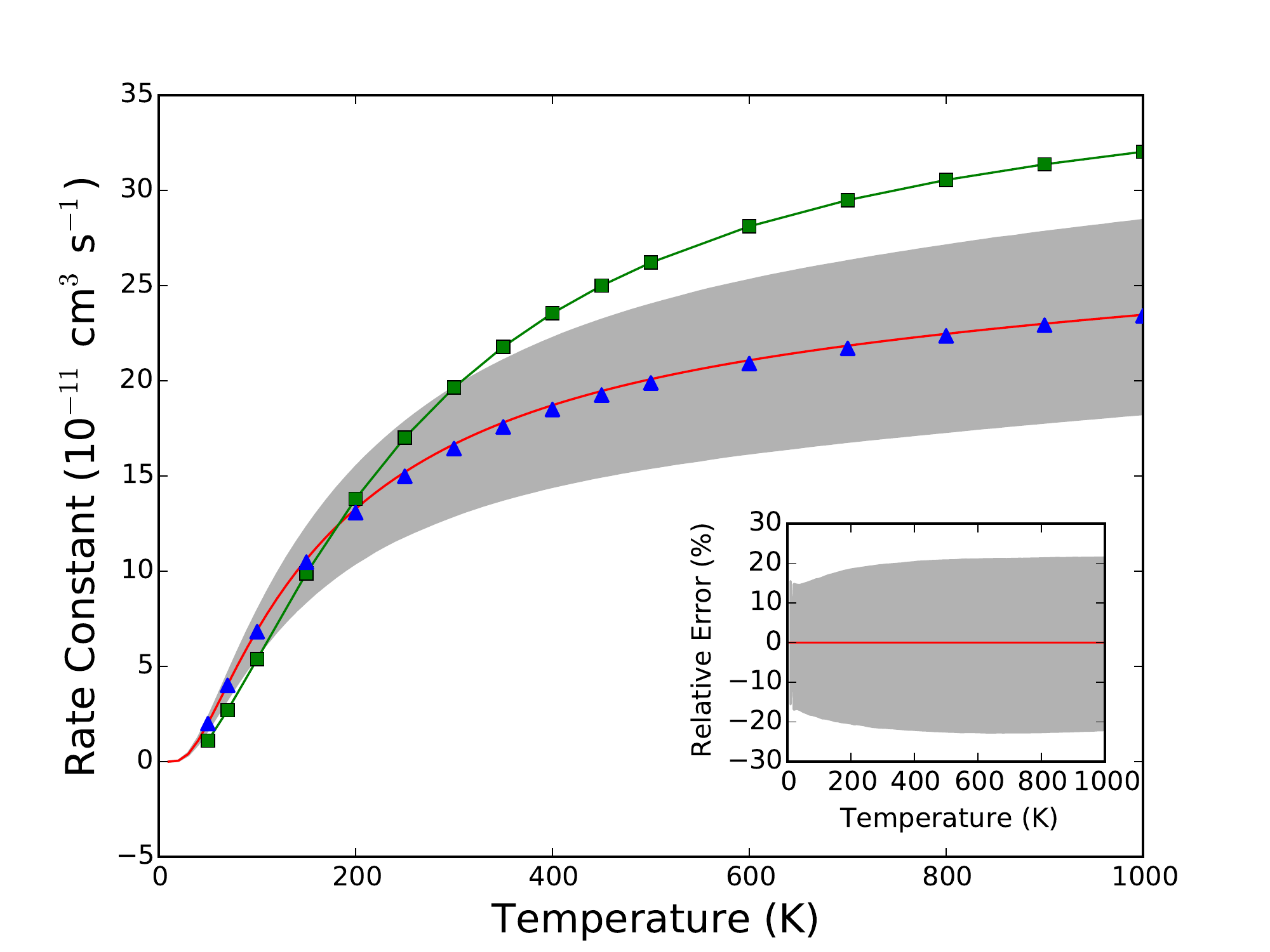}
\includegraphics[scale=0.4]{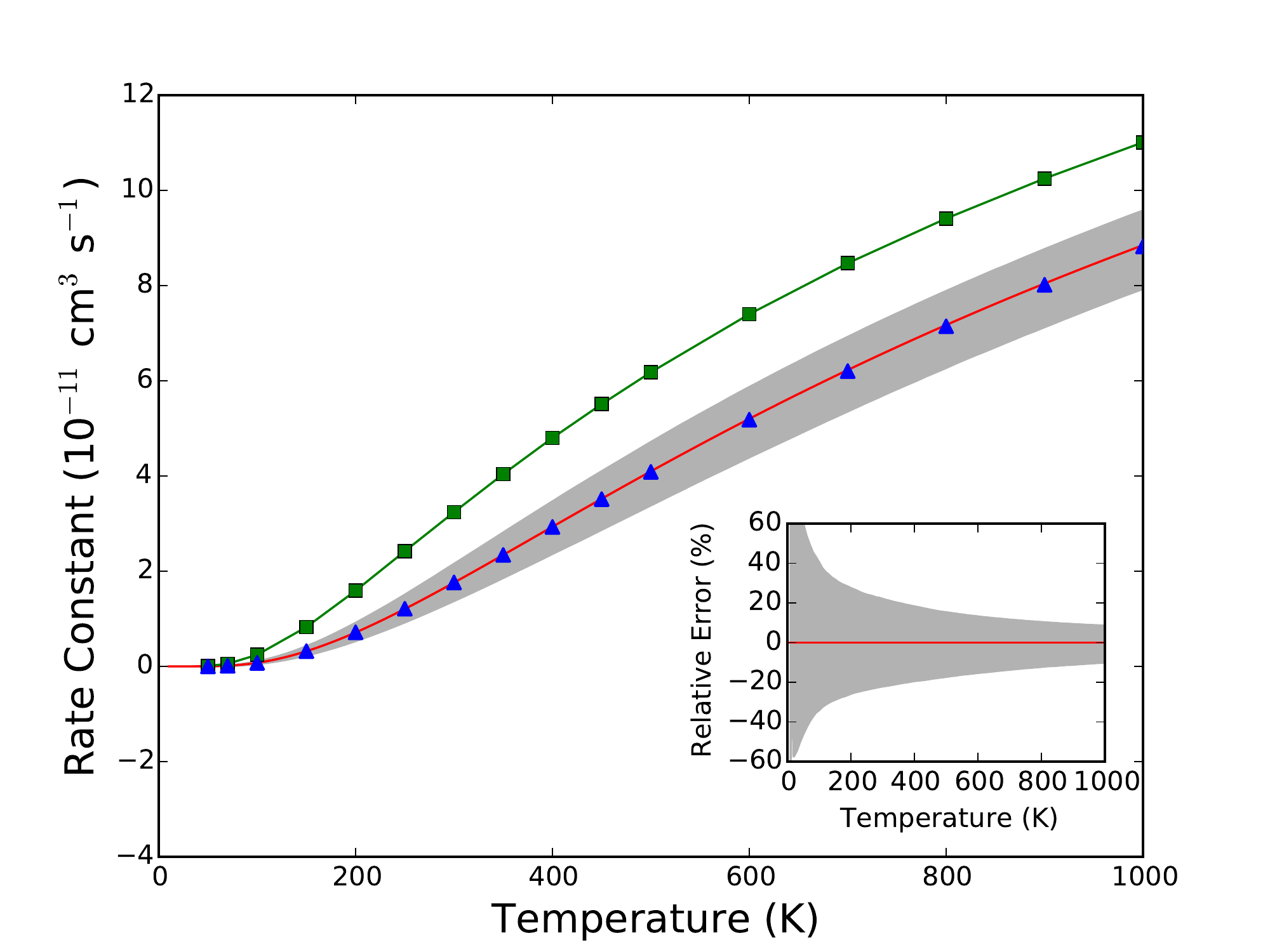}
\includegraphics[scale=0.4]{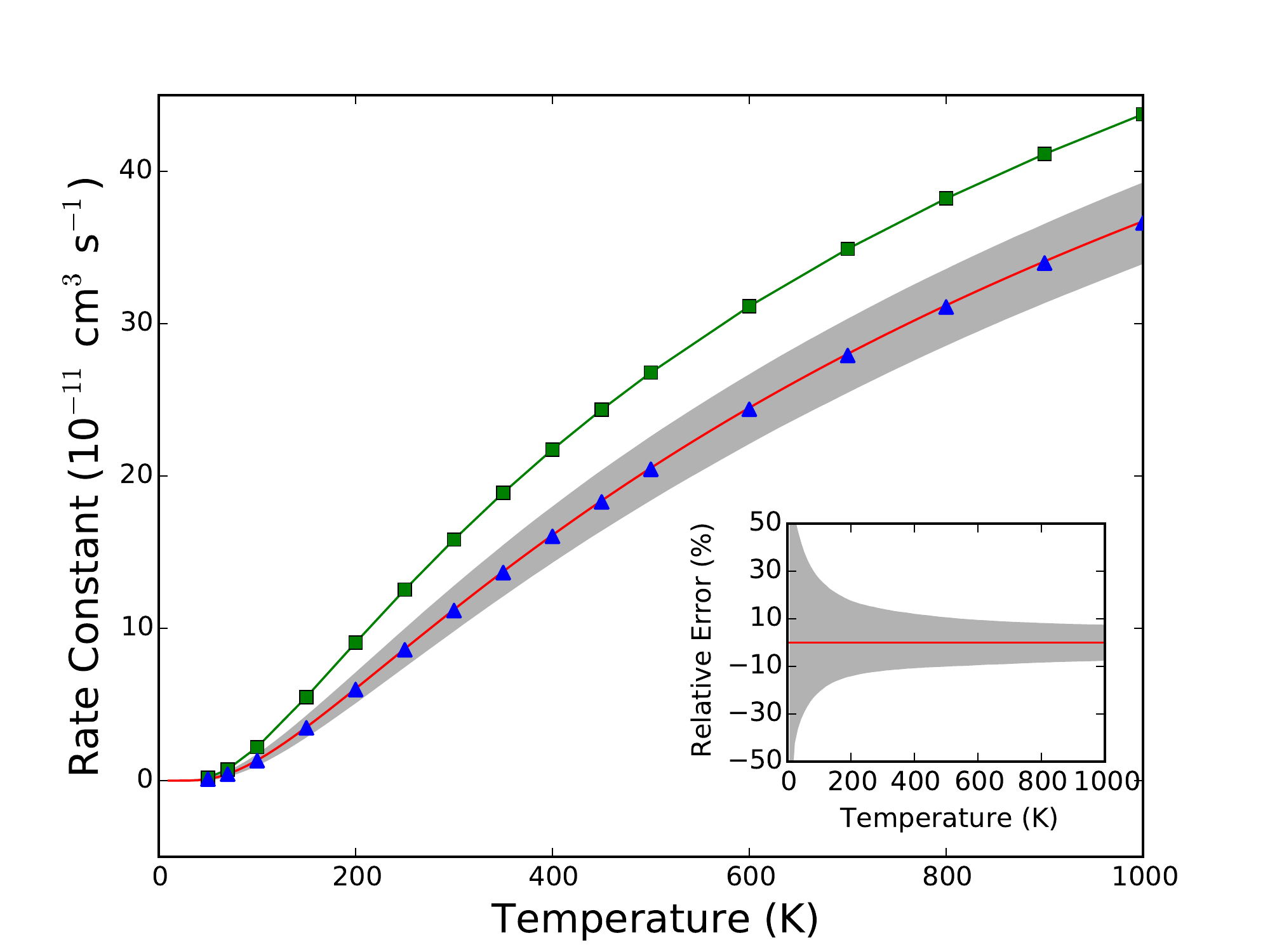}
\caption{Rate constants with error bars for the {\color{black} fine-structure} transitions: {\color{black} $j=1 \rightarrow j' = 0$ (top left), $j=2 \rightarrow j' = 0$ (top right), $j=2 \rightarrow j' = 1$ (bottom).} The red curve is the mean rate constant predicted by the GP model, trained with 2000 points. The blue triangles are the rate constants computed by rigorous coupled channel calculations based on the interaction potentials shown by solid lines in Figure 1.
The trained GP model predicts 90\% of the rates to be within the grey bands, given the $\pm$ 20\% variation in the potentials shown by the grey bands in Figure 1. The insets show the error bars of the rate constants as a percentage deviation from the mean calculated rate constant vs temperature. The curve with green squares shows the results of Abrahamsson et al (2007).}
\end{figure}

\section{Results}

\begin{figure}
\centering
\includegraphics[scale=0.4]{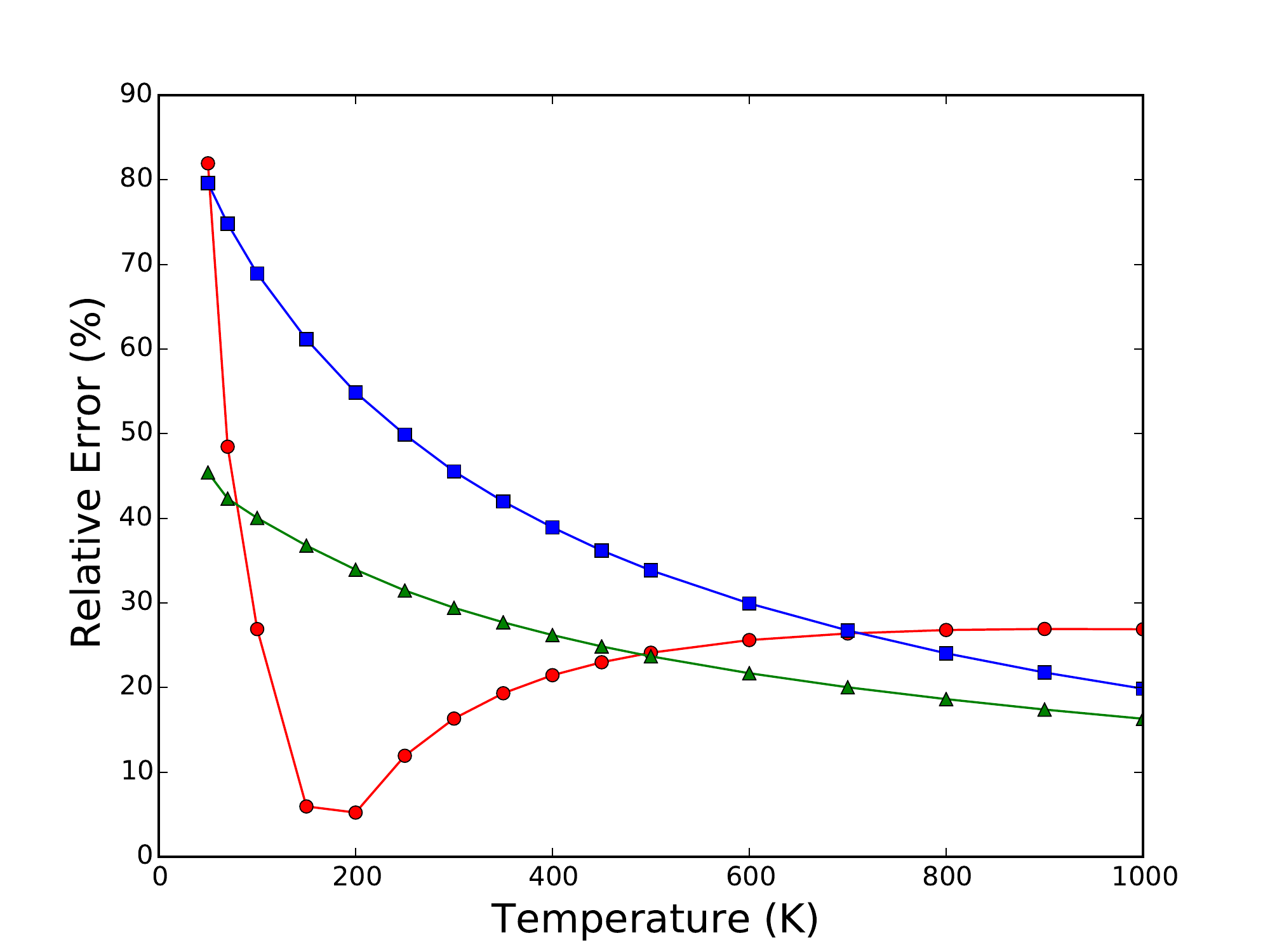}
\caption{Percentage deviation of the rate constants computed here from the results of Abrahamsson et al (2007). The red circles -- {\color{black} the $j=1 \rightarrow j' = 0$ transition; the blue squares -- the $j=2 \rightarrow j' = 0$ transition; the green triangles --the $j=2 \rightarrow j' = 1$ transition.}}
\end{figure}

We begin by illustrating the accuracy and efficiency of GP regression for the prediction of rate constants for the reaction (\ref{reaction}). Figure 2 presents a sample of 200 rate constants for the {\color{black} $j=1 \rightarrow j' = 0$} transition at $T = 300$ K randomly selected in the five-dimensional space of the potential parameters. For each point, we compute the rate constant with the coupled channel method and with Eq. (\ref{mean-prediction}) based on three different numbers of training points: $Q=10$, $Q=100$ and $Q=1000$. The results illustrate that 1000 scattering calculations produce an accurate GP model giving the five-dimensional dependence of the rate constants on the potential parameters. 
Note that the efficiency of GP regression is known to improve with increasing dimensionality of the problem, i.e. the total number of points divided by the number of dimensions should decrease with increasing dimensionality (Deisenroth 2010). The results of Figure 2 thus indicate that it is feasible to construct a GP model of rate constants parametrized by as many as 50 parameters. Constructing such a model would require  an iterative computation of the inverse of the correlation matrix of the size less than 10,000 $\times$ 10,000. 
For the following calculations, we use the GP model trained by $Q=2000$ scattering calculations. This number should be contrasted with $10^5$, which would be the number of scattering calculations to construct the global dependence of the rate constants on the five potentials parameters, if the calculations were performed on a grid with 10 points per parameter. 

Figure 3 presents the temperature dependence of the rate constants for {\color{black} fine-structure} excitations with the error bars corresponding to the variations of the interaction potentials described in the previous sections and depicted in Figure 1. These temperature dependences are obtained from the GP model. As an additional check of the accuracy of these results, we computed the rate constants with the coupled channel approach using the potentials of Parlant and Yarkony (1999). The results, shown by blue triangles in Figure 3, are graphically indistinguishable from the means of the uncertainty intervals represented by the red solid lines.

Figure 3 also includes the data of Abrahamsson et al (2007) represented by the line with green squares. As can be seen their data underestimate or overestimate the current calculation by almost a factor of 2 at low temperatures and $\sim 10 - 30$ \% at elevated temperatures. 
To illustrate the discrepancy between the previous calculations and the current results, we plot the difference of the rate constants vs temperature in Figure 4. We have traced the origin of the discrepancy to the incorrect long-range behaviour of the interaction potentials used by Abrahamsson et al (2007) resulting from an extrapolation of the data of Parlant and Yarkony (1999). {\color{black} The $^{2}$$\Sigma$, $^{4}$$\Pi$ and $^{4}$$\Sigma$ potentials of Abrahamsson et al (2007) become degenerate at interatomic separations $R > 6$ bohr, as a result of the matching of the potentials to the long-range analytical form with the same dispersion coefficient. As illustrated in the inset of Figure 1, the three degenerate excited-state potentials used by Abrahamsson et al (2007) have an abnormally large van-der-Waals minimum at long range. 
Recently, Dagdigian et al (2016) demonstrated that these molecular potentials in fact have a very shallow long-range minimum, not exceeding 27 cm$^{-1}$.}

For future reference, we present the numerical values of the rate constants for the {\color{black} fine-structure} excitation transitions in Tables I - III. The tables include the data obtained by Launay \& Roueff (1977) and by Abrahamsson et al (2007), as well as the uncertainty intervals shown in Figure 3. 

Since GP regression produces a global dependence of rates on the potential energy parameters, it can be used to explore the sensitivity of the rate constants to different interaction potentials. To do this, we perform the variance-based sensitivity analysis as described by multiple authors (Mcrae et al. 1982, Saltelli et al. 1999, Saltelli \& Bolado 1998, Cacuci \& Ionescu-Bujor 2004, Helton \& Davis 2002). This analysis produces five coefficients that quantify the contribution of each underlying parameter to the variance of the rate constants. Figure 5 presents the temperature dependence of these coefficients for the three {\color{black} fine-structure} excitations. 

Surprisingly, we find that the {\color{black} $j=1 \rightarrow j' = 0$} transition is predominantly determined by a single potential of the $^{2}$$\Sigma$ electronic states. For the other transitions, we find that the dynamics of {\color{black} fine-structure} excitations at low temperatures 
is determined by a combination of all four potentials, each with a varying degree of importance. However, at temperatures above 500 K, the rate constants for the {\color{black} $j=2 \rightarrow j' = 1$ and $j = 2 \rightarrow j' = 0$} transitions are much more sensitive to variations of the 
 $^{4}$$\Pi$ potential than to variations in any of the other potentials. We have verified these result by direct coupled channel calculations of the rates as functions of different potential parameters.

The sensitivity analysis is useful for practical purposes. It can be used to guide further refinement of the rate constant calculations. For example, if the rate constants for the {\color{black} $j=1 \rightarrow j' = 0$} transition need to be computed with high precision, our results show that the ab initio calculations must focus on determining the interaction potential for the $^{2}$$\Sigma$ state. In order to refine the rate constants for the {\color{black} $j=2 \rightarrow j' = 1$ and $j = 2 \rightarrow j' = 0$} transitions at high temperatures, the ab initio calculations must refine the $^{4}$$\Pi$ potential. 
On the other hand, the attractive $^2\Pi$ potential is important for the rate constants at low temperatures. Note that moving the position of the minimum or the repulsive wall of this potential does not appear to have a significant effect on the rate constants at any temperature.

\section{Conclusion}
In this work, we have computed the rate constants for the {\color{black} fine-structure} transitions in collisions of O($^3P_j$) with H($^2S$) of importance in astrophysics using the most accurate potentials available to date. 
We have developed an approach to analyze the variation of the rate constants upon modulation of the interaction potentials for collision systems with complex potentials. 
The {\color{black} fine-structure} transitions in  O($^3P_j$) + H collisions represent an example of non-adiabatic dynamics determined by multiple adiabatic interaction potentials. 
The method described here can be used to obtain information on the sensitivity of the rate constants to each of the underlying potentials and to compute error bars of the rate constants corresponding to a certain variation of 
each potential. In this work, we have assumed that the potentials are known to better than 20 \% and computed the distributions of rate constants corresponding to the 20 \% variation of all four potentials determining the collision dynamics. The approach presented here can be easily extended to systems with more potentials. 

We note that the exact uncertainty of the molecular potentials is difficult to assess. In a recent work, Faure et al (2016) computed the interaction potential for the H$_2$ + CO molecular system with extremely high precision. The authors estimate the uncertainty of the ab initio calculations to be within 1 \%. To the best of our knowledge, the calculation of Faure et al (2016) is one of the most accurate quantum chemistry calculations to date. To compute the interaction potentials for the OH molecular systems, Parlant \& Yarkony (1999) used a state-of-the-art multi-reference configuration interaction method. However, the open-shell electronic structure of the OH molecule makes these calculations difficult. 
The calculations of Parlant \& Yarkony (1999) should thus be expected to be less accurate than those of Faure et al (2016). However, given that the ab initio calculations can produce results accurate to close to 1 \%, we believe that the error of 20 \% adopted here for the OH molecular potentials is an overestimate. Our results show that the 20 \% variation of the potentials affects the rate constants by less than 50 \% at low temperatures and less than 20 \% at elevated temperatures. These errors are generally acceptable for models used in astrophysics. Further refinement of the O($^3P$) + H($^2S$) scattering rates, if necessary, must focus on assessing and improving the accuracy of the molecular interaction potentials, guided by the results of Figure 5.

We thank Professor Paul Dagdigian for the discussions that have led us to discover the issue with the potentials used by Abrahamsson et al (2007). The work was supported by NSERC of Canada.

 \begin{figure}
\centering
\includegraphics[scale=0.4]{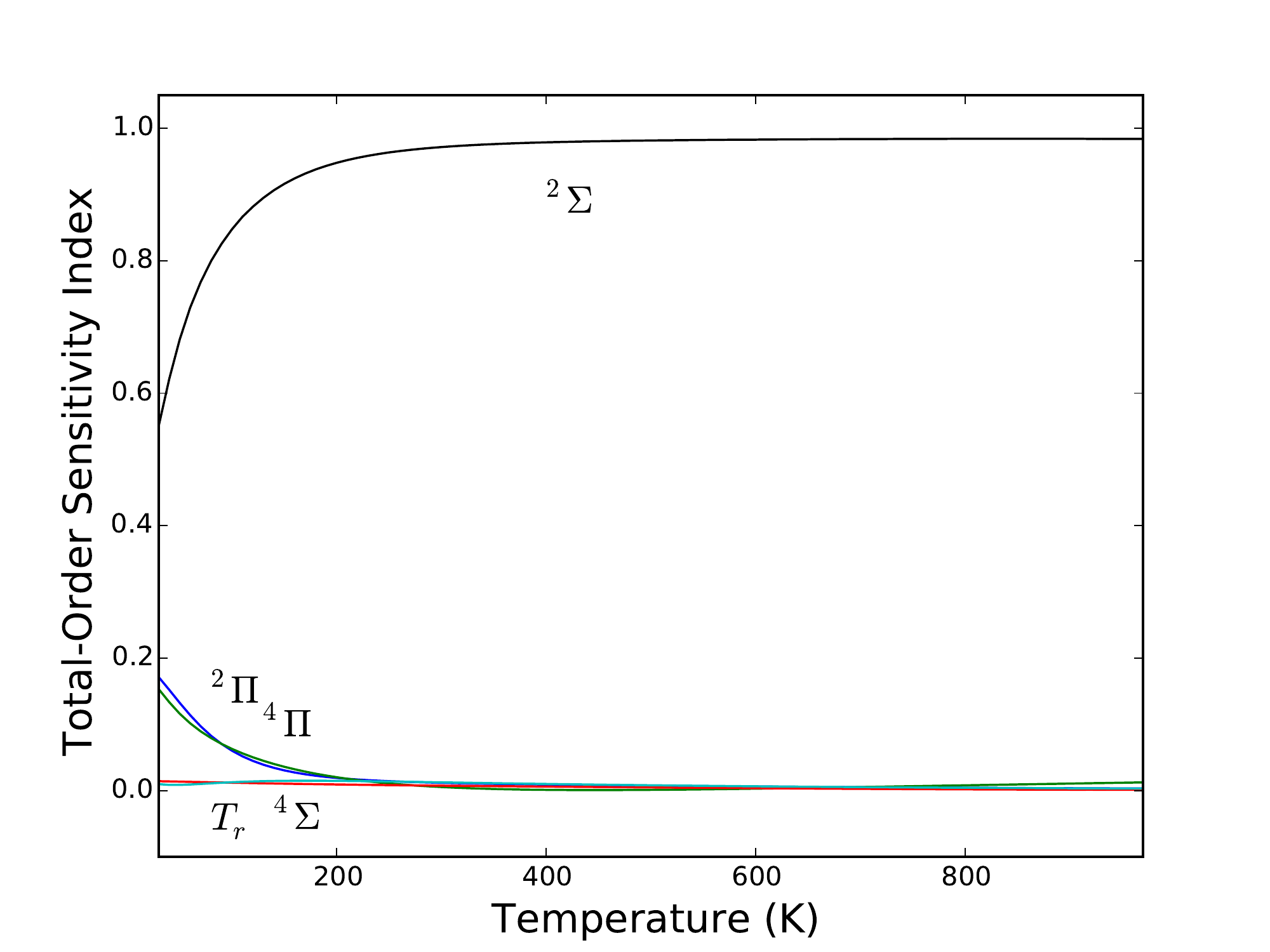}
\includegraphics[scale=0.4]{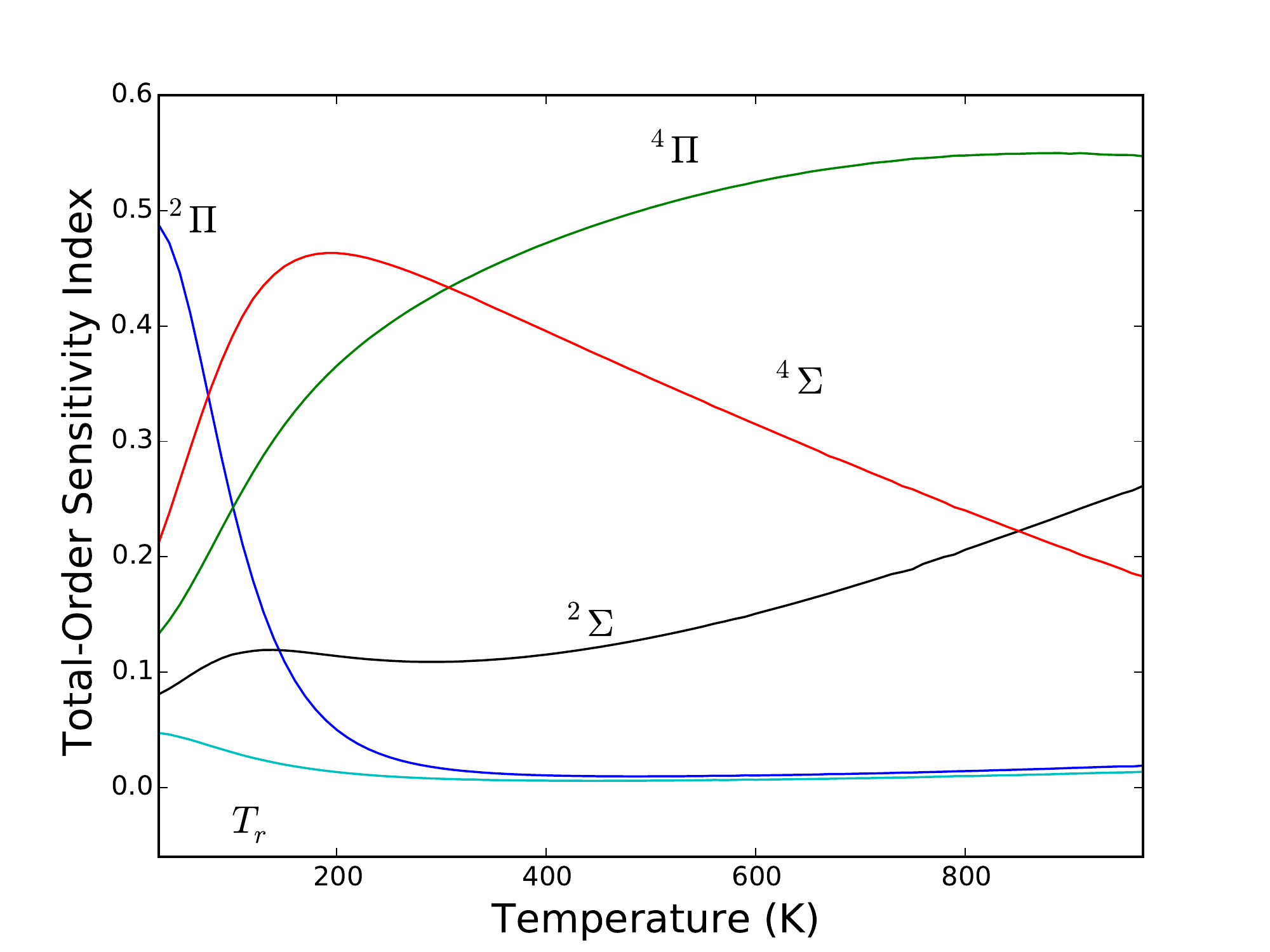}
\includegraphics[scale=0.4]{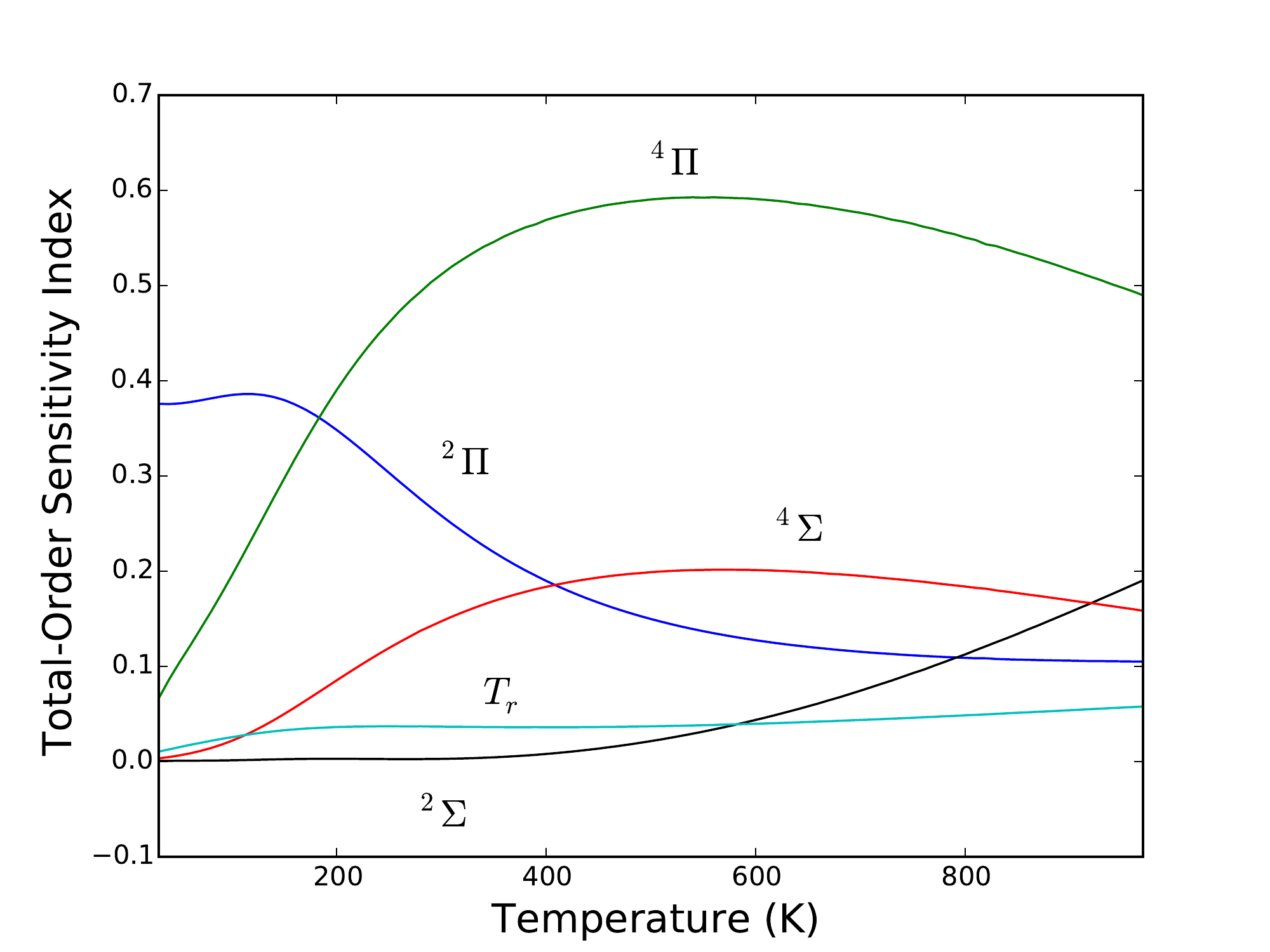}
\caption{The sensitivity of the rate constants for the {\color{black} $j=1 \rightarrow j' = 0$ transitions (top left), $j=2 \rightarrow j' = 0$ transition (top right) and $j=2 \rightarrow j' = 1$ transition (bottom)} to the potential energy parameters. The parameter $T_r$ translates the potential X$^2\Pi$ along {\color{black} the interatomic distance} by $\pm 10$ \%.} 
\end{figure}

\clearpage
\newpage

\setcounter{table}{0}
\begin{table}[h!]
\renewcommand{\thetable}{\arabic{table}}
\centering
\caption{Rate constant for the {\color{black} $j=1 \rightarrow j' = 0$} transition (in units of 10$^{-11}$ cm$^{3}$ s$^{-1}$)} \label{tab:decimal}
\begin{tabular}{ccccc}
\tablewidth{0pt}
\hline
\hline
\multicolumn1c{T (K)} & \multicolumn1c{LR '77} & \multicolumn1c{Abr. '07} & \multicolumn1c{This Work} & \multicolumn1c{Middle 90\%} \\
50 & 0.39 & 1.10  &  2.00 &  1.65-2.30  \\
70 & ... & 2.70  &  4.00  &  3.28-4.62  \\
100 & 1.33 &  5.38  &   6.83  &  5.55-7.92 \\
150 & 2.13 & 9.89  &   10.48 & 8.41-12.30  \\
200 & 2.81 & 13.80   &   13.08 & 10.42-15.49 \\
250 & 3.39  & 17.02   &     14.99 & 11.84-17.84  \\
300 & 3.92  &  19.65   &     16.44 & 12.91-19.65 \\
350 &  4.40 &   21.79   &    17.58 &  13.76-21.07 \\
400 & 4.85  &   23.55   &    18.50 &  14.43-22.24 \\
450 &  5.28  &  25.00   &    19.25 &  14.97-23.21 \\
500 & 5.69  &   26.21   &    19.89 &  15.43-24.01 \\
600 &  6.48  &  28.11   &   20.91 &   16.19-25.29 \\
700 &  7.21  &  29.49   &   21.71 &   16.80-26.28 \\
800 &  7.91  &  30.55   &    22.36 & 17.32-27.11 \\
900 &  8.58  &  31.37   &    22.92 & 17.80-27.82 \\
1000 & 9.22  & 32.03   &    23.42 & 18.25-28.45  \\
\hline
\end{tabular}
\end{table}

\setcounter{table}{1}
\begin{table}[h!]
\renewcommand{\thetable}{\arabic{table}}
\centering
\caption{Rate constant for the {\color{black} $j=2 \rightarrow j' = 0$} transition (in units of 10$^{-11}$ cm$^{3}$ s$^{-1}$)} \label{tab:decimal}
\begin{tabular}{ccccc}
\tablewidth{0pt}
\hline
\hline
\multicolumn1c{T (K)} & \multicolumn1c{LR '77} & \multicolumn1c{Abr. '07} & \multicolumn1c{This Work} & \multicolumn1c{Middle 90\%} \\
50 & ... &  0.0080  &   0.0016 & 0.0009-0.0026  \\
70 & ... & 0.0548  &  0.0138 & 0.0083-0.0206  \\
100 & ... &  0.243  &   0.0755 & 0.0498-0.106  \\
150 & 0.13 &  0.827  &   0.321 & 0.229-0.423 \\
200 & 0.29 & 1.59   &   0.717 & 0.533-0.915 \\
250 & 0.49 &  2.42   &     1.21 &  0.924-1.51  \\
300 & 0.71 &   3.24   &     1.76 & 1.37-2.15 \\
350 & 0.94 &   4.04   &    2.34 & 1.86-2.81 \\
400 & 1.18 &   4.80   &    2.93 & 2.36-3.47 \\
450 & 1.42 &   5.51   &    3.52 & 2.87-4.10 \\
500 & 1.66 &   6.18   &    4.09 & 3.38-4.71 \\
600 & 2.13 &   7.40   &   5.18  & 4.39-5.87 \\
700 & 2.57 &  8.47   &   6.20 & 5.36-6.92 \\
800 & 3.00 &   9.41   &    7.15 & 6.27-7.89 \\
900 & 3.41 &   10.25   &    8.02 & 7.12-8.77 \\
1000 & 3.80 & 11.01   &    8.82 & 7.93-9.57 \\
\hline
\end{tabular}
\end{table}

\setcounter{table}{2}
\begin{table}[h!]
\renewcommand{\thetable}{\arabic{table}}
\centering
\caption{Rate constant for the {\color{black} $j=2 \rightarrow j' = 1$} transition (in units of 10$^{-11}$ cm$^{3}$ s$^{-1}$)} \label{tab:decimal}
\begin{tabular}{ccccc}
\tablewidth{0pt}
\hline
\hline
\multicolumn1c{T (K)} & \multicolumn1c{LR '77} & \multicolumn1c{Abr. '07} & \multicolumn1c{This Work} & \multicolumn1c{Middle 90\%} \\
50 & ... &  0.189  &  0.103 &  0.0732-0.142  \\
70 & ... & 0.740  &  0.427 & 0.322-0.563  \\
100 & 0.56 &  2.20  &   1.32 & 1.05-1.66 \\
150 & 1.66 &  5.49  &   3.47 & 2.92-4.19 \\
200 & 2.98 &  9.06   &   5.99 & 5.17-7.01 \\
250 & 4.37 &  12.54   &     8.59 & 7.54-9.90  \\
300 & 5.74 &   15.83   &     11.17 & 9.89-12.70 \\
350 & 7.06 &   18.90   &    13.66 & 12.19-15.38 \\
400 & 8.33  &  21.74   &    16.04 & 14.39-17.91 \\
450 & 9.54  &  24.36   &    18.30 & 16.49-20.30 \\
500 & 10.69 &   26.79   &    20.44 & 18.48-22.52 \\
600 & 12.80 &   31.14   &   24.39  & 22.19-26.59 \\
700 & 14.80 &    34.91   &   27.92  & 25.54-30.24 \\
800 & 16.60 &    38.22   &    31.10 & 28.63-33.51 \\
900 & 18.30  &  41.14   &    33.98 & 31.43-36.48 \\
1000 & 19.90 &   43.75   &    36.61 & 34.00-39.19 \\
\hline
\end{tabular}
\end{table}

\clearpage
\newpage


\end{document}